\documentclass[journal=jacsat,manuscript=article]{achemso}
\usepackage{pdfpages}
\UseRawInputEncoding

\usepackage{braket}
\usepackage{bbold}
\usepackage{soul}
\usepackage[version=3]{mhchem} 

\author{Yotam Wolf}
\author{Yizhou Liu}
\author{Jiewen Xiao}
\affiliation{Department of Condensed Matter Physics, Weizmann Institute of Science, Rehovot 7610001, Israel}
\author{ Noejung Park}
\affiliation{Department of Physics, Ulsan National Institute of Science and Technology (UNIST), Ulsan, 44919
Republic of Korea}
\author{Binghai Yan}
\email{binghai.yan@weizmann.ac.il}
\affiliation{Department of Condensed Matter Physics, Weizmann Institute of Science, Rehovot 7610001, Israel}

\title[An \textsf{achemso} demo]
  {Unusual Spin Polarization in the Chirality Induced Spin Selectivity}


\makeatletter
\newcommand*{\forcekeywords}{
  \acs@keywords@print
  \let\acs@keywords@print\relax
}
\makeatother

\begin{document}

\begin{tocentry}

\includegraphics[width=7cm]{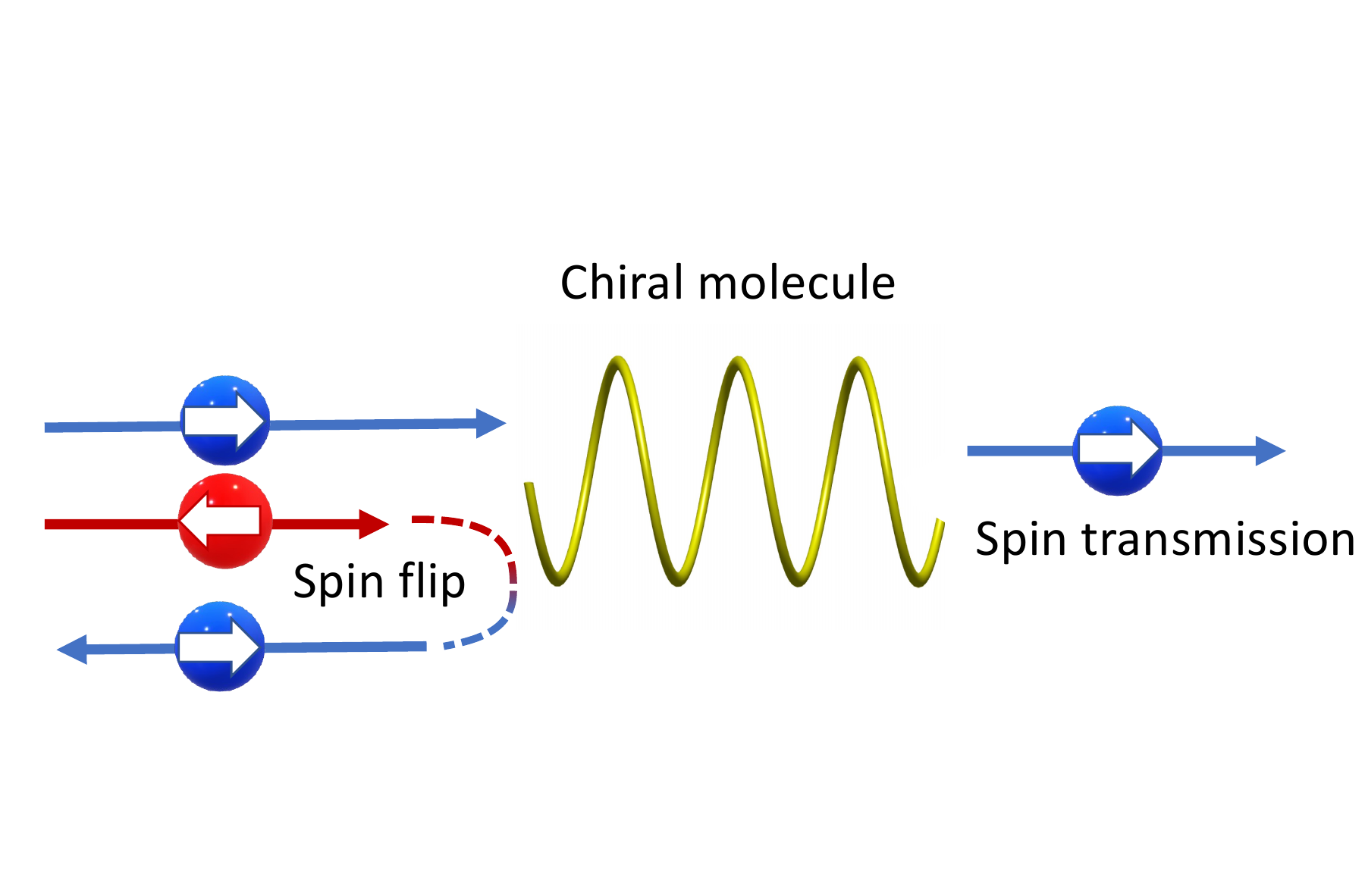}

\end{tocentry}

\begin{abstract}
  Chirality-induced spin selectivity (CISS) refers to the fact that electrons get spin polarized after passing through chiral molecules in a nanoscale transport device or in photoemission experiments. In CISS, chiral molecules are commonly believed to be a spin filter through which one favored spin transmits and the opposite spin gets reflected, i.e., transmitted and reflected electrons exhibit opposite spin polarization.
In this work, we point out that such a spin filter scenario contradicts the principle that equilibrium spin current must vanish. Instead, we find that both transmitted and reflected electrons present the same type spin polarization, which is actually ubiquitous for a two-terminal device. More accurately, chiral molecules play the role of a spin polarizer rather than a spin filter. The direction of spin polarization is determined by the molecule chirality and the electron incident direction. And the magnitude of spin polarization replies on local spin-orbit coupling in the device. Our work brings a deeper understanding on CISS and interprets recent experiments, for example, the CISS-driven anomalous Hall effect. 
\end{abstract}

\section{Introduction}
Chirality induced spin selectivity (CISS) is a fascinating effect where electrons get spin polarized after propagating through chiral organic molecules like DNA~\cite{Ray1999,Gohler2011,Naaman2012,Naaman2019} and inorganic materials such as oxides\cite{mollers2022spin,ghosh2019controlling,ghosh2020increasing} and perovskites ~\cite{lu2019spin,wan2022anomalous,zhang2021great,kim2021chiral} .
CISS reveals an intriguing relation between the structural chirality and the electron spin or orbital\cite{liu2021chirality,Gersten2013} , and is promising to design spintronic devices using chiral molecules\cite{yang2021chiral} , realize enantiomers separation\cite{Banerjee2018} , and study the spin-selective biological process~\cite{Naaman2012,Naaman2019} .
Despite the debate on the microscopic mechanisms (see Ref. \cite{evers2021theory} and references therein), the chiral molecule is widely regarded a spin filter \cite{yeganeh2009chiral,gutierrez2012spin,maslyuk2018enhanced,dalum2019theory} .

The CISS spin filter represents that the chiral molecule exhibits a selected transmission rate in one spin channel compared to the opposite spin, in which the preferred spin depends on the chirality. The spin filter was presumed to induce opposite spin polarization in transmitted and reflected electrons (see Fig. \ref{fig:spinfilter}a). This scenario was further generalized to argue that a chiral molecule exhibits transient, opposite spin polarization at two ends in the charge displacement process (see Ref. \cite{Naaman2019} for review).  In literature, the spin filter is frequently adopted to rationalize CISS experiments such as the magnetoresistance~\cite{Xie2011,Kettner2015,Naaman2015,Kiran2016,Al2018,Varade2018,Liu2020,Lu2019,Stemer2020,Ghosh2020, lu2020highly,mondal2021spin,lu2021highly} , 
anomalous Hall effect (AHE) \cite{smolinsky2019electric,mishra2020b} ,
and the selected chiral adsorption~\cite{Banerjee2018,Tassinari2019} . 
Such a spin filter scenario is plausibly based on an elusive argument that the total spin density remains zero to preserve the net spin polarization. However, it is established that spin polarization is not necessarily conserved in transport by earlier studies on spintronics, for example, the Rasba-Edelstein effect where the current leads to net spin polarization in a nonmangetic material
\cite{edelstein1990spin,sanchez2013spin}. 
Therefore, this well-accepted model deserves more examination.

In this work, we point out that the present spin filter picture is incompatible with the prohibition of equilibrium spin current or the general time-reversal symmetry analysis.
We prove that both transmitted and reflected electrons present the same type of spin polarization in a two-terminal CISS device based on the unitarity of the scattering matrix. Chiral molecules play the role of a spin polarizer (Fig.~\ref{fig:spinfilter}d) rather than a spin filter, because they polarize all scattered (transmitted and reflected) electrons to the same direction, which relies on the molecule chirality and electron incident direction. 
The spin polarizer picture provides further understandings on CISS, especially on the CISS-driven anomalous Hall effect and the transient spin polarization of chiral molecules in dynamical chemical processes. 
The scope of preset work is to understand the current-induced spin polarization from a nonmagnetic CISS device. We ignore the case involving magnetic electrodes in which CISS-induced magnetoresistance is commonly measured and exhibits essential features beyond the spin polarization \cite{Dalum2019,Yang2020,xiao2022magnetochiral} .

\section{Results and discussion}

\subsection{Prohibition of equilibrium spin current}

\begin{figure}
\includegraphics[width=15cm]{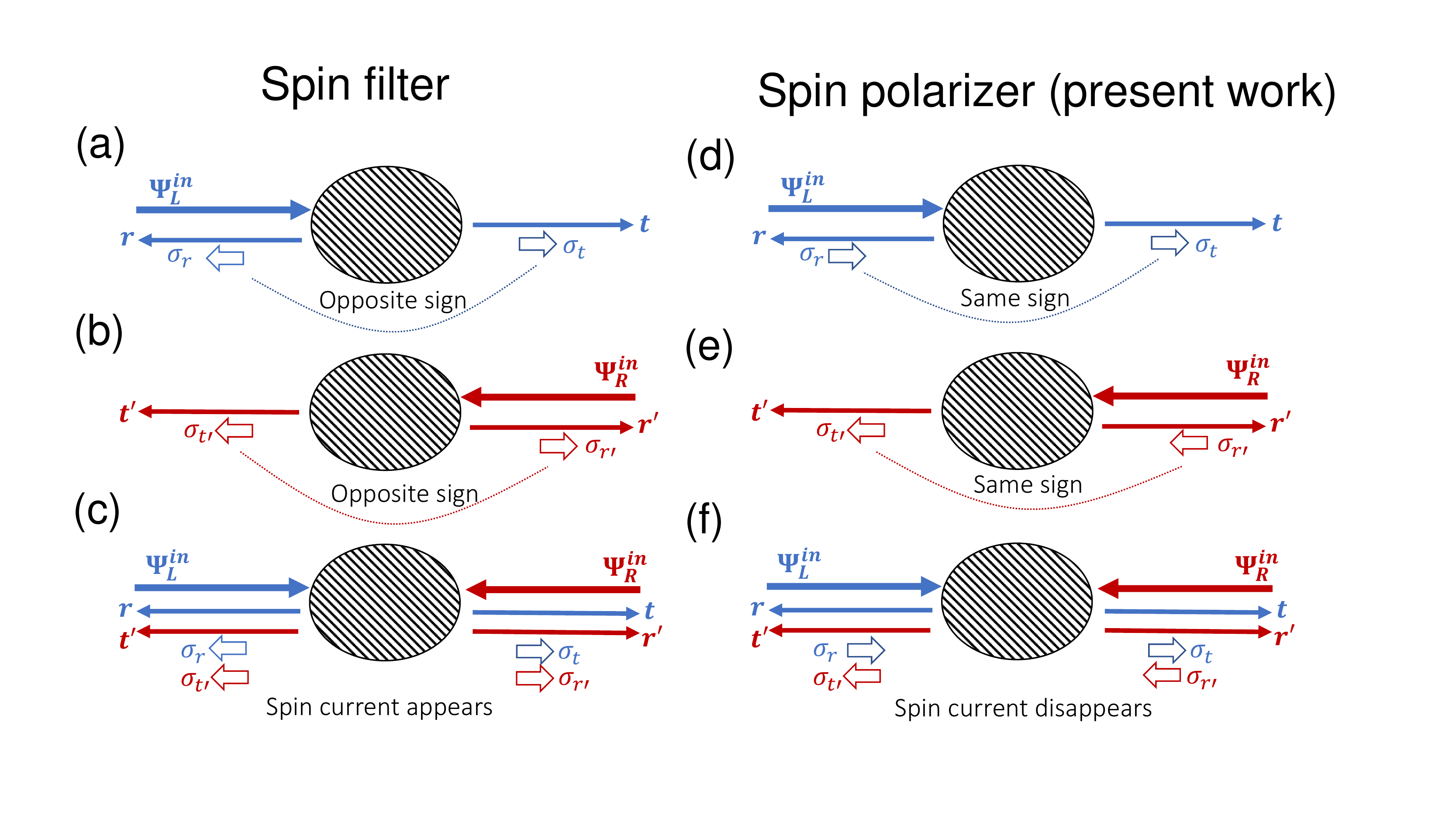}
\caption{\label{fig:spinfilter} Schematics of transmission and reflection in a two-terminal device. 
The incident wave from the left (right), $\Psi^{in}_{L}$ ($\Psi^{in}_{R}$), gets scattered with transmission rate $t$ ($t^\prime$) and reflection rate $r$ ($r^\prime$). The transmission and reflection exhibit spin conductance, $\sigma_t$ ($\sigma_{t^\prime}$) and $\sigma_r$ ($\sigma_{r^\prime}$), respectively. 
The spin filter requires that $\sigma_t$ ($\sigma_{t^\prime}$) and $\sigma_r$ ($\sigma_{r^\prime}$) have opposite signs in (a) [(b)]. 
 In (c) electrons come from the left ($\Psi_L^{in}$) and right ($\Psi_R^{in}$) equally, which represents the equilibrium state. Then, the left ($r,t'$) and right ($t,r'$) scattered waves carry opposite spins. 
Therefore, the equilibrium spin current emerges in (c), which is unphysical.
In contrast, the spin polarizer leads to the same sign in  $\sigma_t$ ($\sigma_{t^\prime}$) and $\sigma_r$ ($\sigma_{r^\prime}$) in (d) [(e)]. Then, the equilibrium  spin current can be avoided in (f) if Eqs. \ref{eq:spin1} and \ref{eq:spin2} hold. }
\end{figure}

We will discuss a generic two terminal device with nonmagnetic electrodes. It is established that the equilibrium spin current is strictly forbidden between two terminals because of the time reversal symmetry \cite{ESC} . In the quantum scattering problem, electrons income from the left or right to the center region and are transmitted ($t$ from the left, $t'$ from the right) or reflected ($r$ from the left, $r'$ from the right), as shown in Fig. \ref{fig:spinfilter}. We denote the spin conductance of a scattering state as ${\sigma}_i = i_{\uparrow\uparrow}-i_{\downarrow\downarrow}$ where 
$i=t,t',r,r'$ is a density matrix. We note that the spin conductance is different from the ratio of spin polarization.

As shown in Fig.~\ref{fig:spinfilter}, we denote by ``spin filter" the scenario that transmission and reflection have opposite spin polarization, and by``spin polarizer" the scenario that they have the same sign.
In a CISS device, the 180$^\circ$ rotation of the molecule gives rise to the same sign of spin polarization in transmission because the rotation does not change the chirality. For given chirality, the polarization direction of $\sigma_t$ is locked to the transmission direction in a parallel or antiparallel way. Thus, $\sigma_t$ is expected to be opposite to $\sigma_{t'}$ in sign because of opposite transmission directions. 
In equilibrium, electrons are equally incident from and scattered to the left and right. The spin filter presents an equilibrium spin current because left-polarized ($\sigma_r,\sigma_{t'}$) and right-polarized ($\sigma_t,\sigma_{r'}$) spins simultaneously move to the left and right, respectively, in Fig.~\ref{fig:spinfilter}c.

In contrast, the spin polarizer can avoid the spin current at equilibrium as long as the spin {conductance} satisfies the condition (Fig.~\ref{fig:spinfilter}f),
\begin{align} 
\sigma_t + \sigma_{r^\prime} = &  0 \label{eq:spin1}\\
\sigma_r + \sigma_{t^\prime} = &  0 \label{eq:spin2}
\end{align}
We will prove that this condition is guaranteed by the unitary property of the scattering matrix in the following.

\subsection{Scattering matrix and spin polarization}

In the two-terminal device, we define the scattering matrix in the usual notation, as the matrix relating the incoming waves, $\Psi^{in}_L$ and $\Psi^{in}_{R}$, to the outgoing waves, $\Psi^{out}_L$ and $\Psi^{out}_R$:
\begin{equation} \label{scattering matrix}
\begin{pmatrix}
\Psi^{out}_L \\
\Psi^{out}_{R}
\end{pmatrix}  =
S 
\begin{pmatrix}
\Psi^{in}_L \\
\Psi^{in}_{R}
\end{pmatrix} 
= \begin{pmatrix}
r & t'\\
t & r'
\end{pmatrix} 
\begin{pmatrix}
\Psi^{in}_L \\
\Psi^{in}_{R}
\end{pmatrix} 
\end{equation}
The unitary property $S S^{\dagger}=\mathbb{1}$ of 
the $S$-matrix leads to,
\begin{equation} \label{eq:unitarity1}
    rr^{\dagger}+t't'^{\dagger} = \mathbb{1}
\end{equation}
\begin{equation} \label{eq:unitarity2}
    r'r'^{\dagger}+tt^{\dagger} = \mathbb{1}
\end{equation}

$\Psi^{in}_L$ and $\Psi^{in}_R$ are incoming wave functions from the left and right leads, respectively. By assuming $N$ independent channels in each lead, $\Psi^{in}_{L,R}$ have $2N$ dimensions because of the spin degeneracy. The density matrix $\rho_i$ of the scattered wave ($i$) can be expressed as, 
\begin{equation}\label{eq:densitymatrix}
    \rho_r = rr^{\dagger} , \rho_t = tt^{\dagger}, \\
    \rho_{r'} = r'r'^{\dagger} , \rho_{t'} = t't'^{\dagger}, 
\end{equation}

It is convenient to calculate the spin conductance of a specific state ($i$) using the corresponding density matrix ($\rho_i$). 
Denoting the transport axis as the $z$ direction, the spin conductance is 
\begin{equation}\label{eq:sigmaz}
    \sigma_i \equiv \braket{\sigma_z}_{i}=Tr[\sigma_z \rho_i]
\end{equation}
Combing Eqs.~\ref{eq:unitarity1},\ref{eq:unitarity2},\ref{eq:densitymatrix} and \ref{eq:sigmaz}, we can derive Eqs. \ref{eq:spin1} and \ref{eq:spin2}, i.e., the condition to avoid equilibrium spin current.
We must stress that Eqs. \ref{eq:spin1} and \ref{eq:spin2} are generic for any two-terminal devices with nonmagnetic leads, as long as the unitarity of the $S$-matrix holds.

For a symmetric CISS device, the two-fold rotation symmetry requires $\sigma_t = -\sigma_{t'}$. From Eqs. \ref{eq:spin1} and \ref{eq:spin2} we obtain, 
\begin{equation} \label{eq:spinC2}
    {\sigma}_{t} = {\sigma}_r = - {\sigma}_{t'} = - {\sigma}_{r'}.
\end{equation}
A similar relation to Eq.~\ref{eq:spinC2} was also derived based on the Onsager's reciprocal relation in Refs.~\cite{Yang2019,Yang2020}  and via the unitarity of the scattering matrix in Ref.~\cite{Utsumi2020}, all with the assumption of a two-fold rotation symmetry.
{In a general CISS devices beyond the two-fold symmetry constraint, we relax the above conditions to $\sigma_t$ and $\sigma_{t'}$ having opposite signs. From Eqs.~\ref{eq:spin1} amd \ref{eq:spin2}, we obtain that transmission and reflection have the same sign in spin conductance (thus also the same sign of spin polarization),}
\begin{equation} \label{eq:spin-general}
    {\sigma_t}{\sigma_r} > 0 , ~~~~
    {\sigma_{t'}}  {\sigma_{r'}}>0.
\end{equation}
These results support the spin polarizer scenario rather than the spin filter. 

The spin polarizer mechanism requires a spin flipping process upon reflection (also discussed in Refs.~\cite{Yang2019,Yang2020}) such that transmitted and reflected electrons have the same sign of spin polarization. This spin-flipping mechanism can also be understood from a simple time-reversal symmetry argument as illustrated in the supplementary Fig. S1.
The spin polarizer picture remains unchanged under time-reversal operation of the refelecdtion process while the spin filter scenario violates time-reversal symmetry.

More generally, the above results hold for {the conductance of} any operator ${\mathcal{O}}$ that satisfies $Tr[ \mathcal{O}]=0$. {Detailed derivations based on the outcome of the time-reversal symmetry are presented in Appendix A. For example, the orbital angular momentum operator $L$ is such an operator, and angular momentum polarization of transmission and reflection also exhibit the same sign. In summary, the chiral molecule serves as a polarizer, rather than a filter, for spins or orbitals\cite{liu2021chirality} of conducting electrons. }

The above discussion holds for the coherent transport, in which the unitary condition of Eqs. \ref{eq:unitarity1} and \ref{eq:unitarity2} holds. For a dissipating system, we can include dissipation by modifying the unitarity condition, $SS^\dagger = (1-\delta)\mathbb{1} + \epsilon A$, where $\delta$ is a uniform dissipation of all the modes and $\epsilon A$ is a selective dissipation term for the different modes. The matrix $A$ is normalized such that its eigenvalues are no larger than $1$, and $\epsilon$ controls the magnitude of selective dissipation. Following the same process as before, by using Eqs. \ref{eq:densitymatrix} and \ref{eq:sigmaz} and taking the trace on the modified unitary conditions, we get:
\begin{equation}
    \sigma_t = -\sigma_{r'} + \epsilon Tr[\sigma_z A_1]
    \textbf{ }\textbf{ }\textbf{ }\textbf{ }\textbf{ } \sigma_{t'}=-\sigma_r  + \epsilon Tr[\sigma_z A_2]  
\end{equation}
Where $A_1$ is the left upper block of $A$ and $A_2$ is the right lower block. We see that the uniform dissipation $\delta$ does not play a role, only the part that creates selectivity in dissipation of the different modes, and that is bound by $\epsilon Tr[\sigma_z A_1]$ and $ \epsilon Tr[\sigma_z A_2]$. The trace of $\sigma_z A_i$ is no larger than $Tr[|A_i|] \leq 2N$. As there are $2N$ independent modes in each lead, we normalize the spin conductance by $2N$ in order to have $-1\leq\sigma\leq1$. Thus, we get:
\begin{equation}
    |\sigma_t + \sigma_{r'}| < \epsilon  \textbf{ }\textbf{ }\textbf{ }\textbf{ }\textbf{ } |\sigma_{t'} + \sigma_r| < \epsilon
\end{equation}
Assuming $\sigma_t =-\sigma_{t'}$, we get:
\begin{equation}
    |\sigma_t - \sigma_{r}| < \epsilon  \textbf{ }\textbf{ }\textbf{ }\textbf{ }\textbf{ } |\sigma_{t'} - \sigma_{r'}| < \epsilon
\end{equation}
From this, we see that if the spin conductance is larger than the magnitude of the selective dissipation term, then the conductance of transmission and reflection spin have the same sign.
The intuitive explanation to why selective dissipation of modes alters the spin conductance induced by the chirality of the molecule, is that selective dissipation is an orthogonal mechanism that can theoretically polarize spin as well, for example, selective dissipation of spin down states in an unpolarized current leaves the current spin polarized in the up direction.

\subsection{ Quantum transport calculations}

\begin{figure}
\includegraphics[width=12cm]{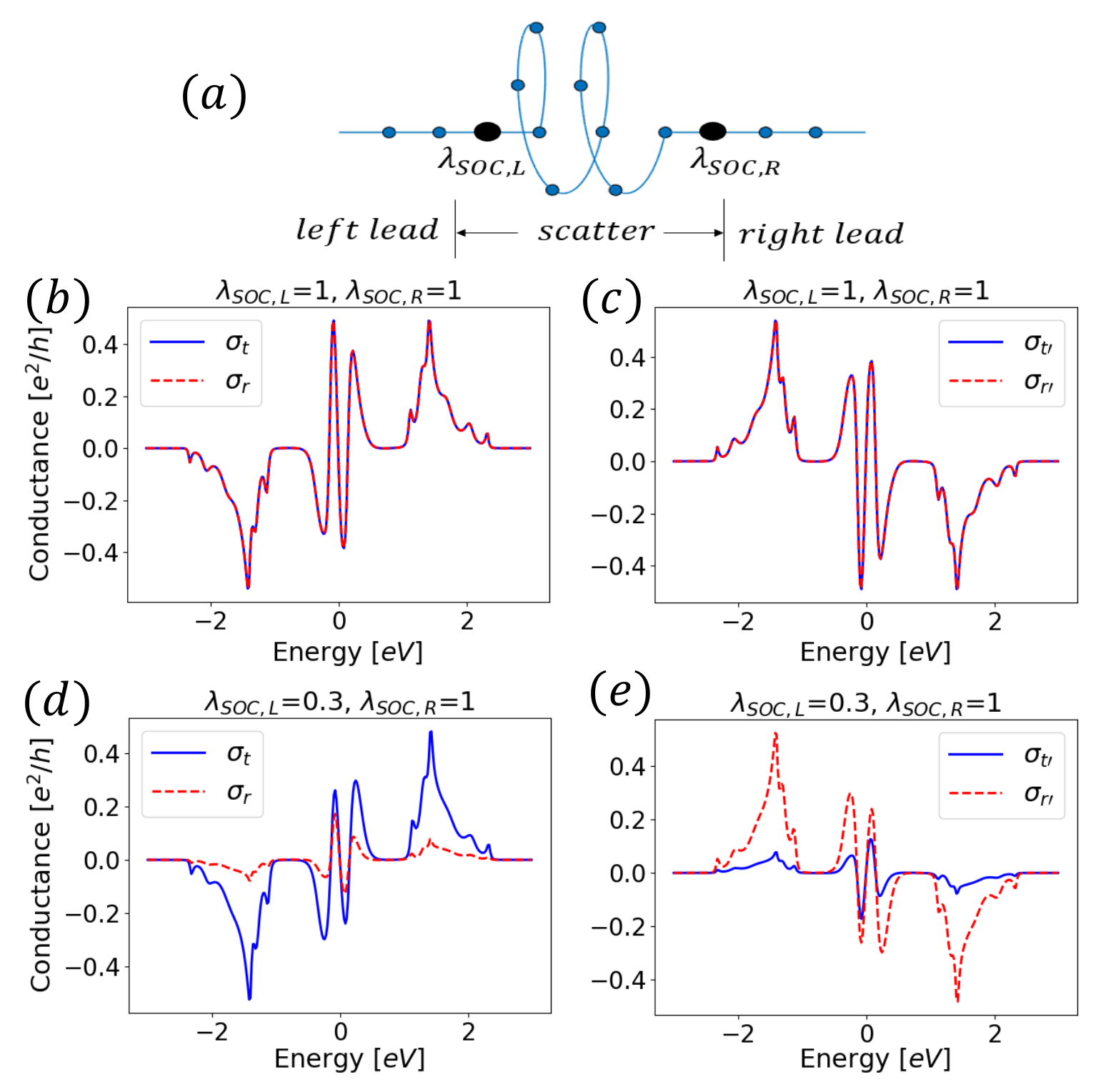}
\caption{\label{fig:calculation1} (a) Generic structure of the CISS device, including the leads and interface. (b)-(c) Calculated spin conductance (${\sigma}_i = i_{\uparrow\uparrow}-i_{\downarrow\downarrow}$, where $i$ is a scattering state) of $t,r,t',r'$ in a $C_2$ symmetric system at different Fermi energies. (d)-(e) The spin conductance in a $C_2$-symmetry-breaking system by differentiating $\lambda_{SOC,L}$ and $\lambda_{SOC,R}$. $\lambda_{SOC,L/R}$ represents the value of spin-orbit coupling at the left/right interface.
}
\end{figure}

To examine the above analytic results, we performed Landauer-B\"{u}tikker quantum transport calculations on two-terminal CISS devices. The device is composed of a helical chain sandwiched between half-infinite linear chains, as illustrated in Fig.~\ref{fig:calculation1}a. Each site has $p_{x,y,z}$ orbitals and two spins. Between sites, we set Slater-Koster type\cite{Slater1954} hopping parameters to nearest neighbors. 
Because the nearest-neighbor $p_{x,y,z}$ hopping parameters depend on the relative atomic positions in a chiral chain, these hopping parameters manifest the chirality by picking up opposite phases for opposite chiralties (see more details in Appendix B). We only set finite spin-orbit coupling (SOC,$\lambda_{SOC}$) at two interface sites, to represent that electrodes exhibit dominantly larger $\lambda_{SOC}$ than the chiral molecule in a CISS device. However, our main conclusions will be independent of whether $\lambda_{SOC}$ comes from the molecule or electrodes. The spin polarization in both terminals is well defined in the calculations because there, we set the SOC to zero. 
All the conductance calculations were performed with Kwant\cite{Kwant} . Parameters of the model and band structure of leads and chiral chain can be found in appendix B.

Figures ~\ref{fig:calculation1}b-c show the case of a symmetric device. The transmission and reflection are calculated for different Fermi energies, which fully agree with Eq.~\ref{eq:spinC2}. After violating the $C_2$ symmetry by setting $\lambda_{SOC}$ which differ between the two interfaces (Fig.~\ref{fig:calculation1}d-e), $\sigma_t$ and $\sigma_r$ ($\sigma_{t'}$ and $\sigma_{r'}$) are not equal in values, but Eqs.\ref{eq:spin1}, \ref{eq:spin2} and \ref{eq:spin-general} are always valid.

Up to this point in the subsection, we presented quantum transport calculations on a specific model to demonstrate the validity of the spin polarizer mechanism. We should point out that the spin polarizer conclusion comes from the unitarity of the scattering matrix and is independent from any details (orbital, spin, or SOC) of the model.

\begin{figure}
\includegraphics[width=12cm]{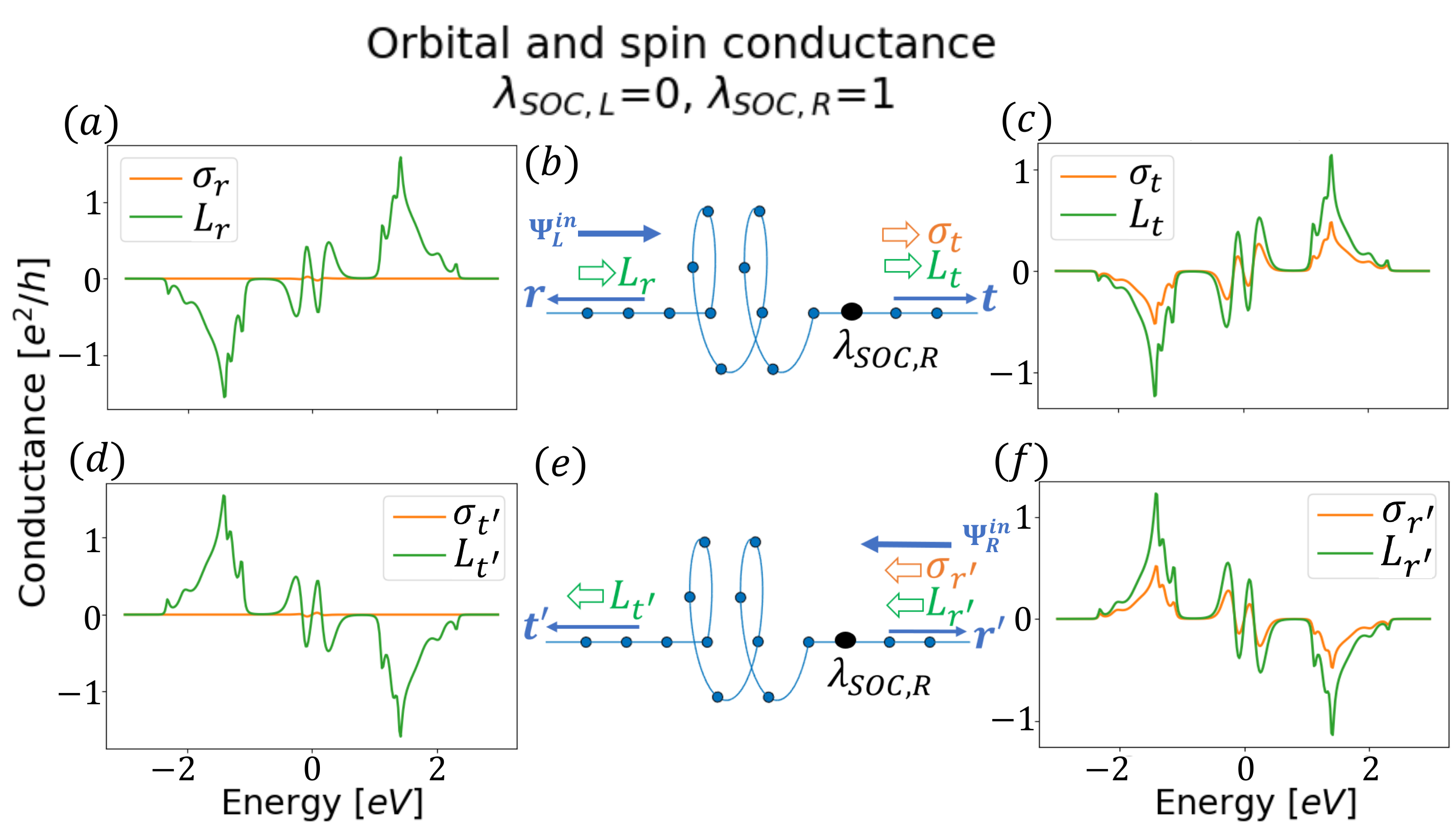}
\caption{\label{fig:orbital-spin} Orbital and spin conductance in a system where $C_2$ symmetry is broken by having SOC only on the right side. (a)-(c) Orbital and spin conductance of $r$ and $t$: Current coming from the left gets orbital polarized along the chain, then, the orbital current is converted to spin current by the SOC atom. (d)-(f) Orbital and spin conductance of $r'$ and $t'$:
Current coming from the right hits the SOC atom before getting orbital polarized, so it passes unaltered into the chain, where it gets orbital polarized, after which it passes to the left lead only orbital polarized.}
\end{figure}

{Figures~\ref{fig:calculation1}d-e already demonstrate sensitive dependence of the spin conductance on SOC. Furthermore, we set an extreme example} with $\lambda_{SOC,L} = 0$ and $\lambda_{SOC,L} = 1$ in Fig.~\ref{fig:orbital-spin} and calculate both the spin and orbital ($L$) conductance here.
One striking feature is that $\sigma_r$ and $\sigma_{t'}$ are diminished in amplitude while $\sigma_{r'}$ and $\sigma_{t}$ are still significant. 
Because the orbital is less sensitive to $\lambda_{SOC}$, the orbital conductance $L_{r,t,r',t'}$ exhibits a large amplitude and satisfies,
\begin{align} 
L_t + L_{r^\prime}  =   0,~~~~~~ L_r + L_{t^\prime}  =   0  \label{eq:orbital1}\\
L_t L_r  >  0 ,~~~~~~~~~ L_{t'} L_{r'}  >  0 \label{eq:orbital3}
\end{align}
One intuitive picture is that electrons get orbital polarized in the chiral molecule and the interface SOC converts the orbital polarization to spin polarization. For example, $\lambda_{SOC,R}$ converts large $L_t$ ($L_{r'}$) to $\sigma_t$ ($\sigma_{r'}$) while there is no $\lambda_{SOC,L}$ to induce $\sigma_r$ ($\sigma_{t'}$) from $L_r$ ($L_{t'}$). In addition, the tiny amplitude of $\sigma_r$ ($\sigma_{t'}$) at some energies is induced by the weak interface orbital-spin conversion due to $\lambda_{SOC,R}$.

In the orbital-spin conversion, we observe a counter-intuitive effect, that the transmitted/reflected spin conductance remains the same when reversing the sign of $\lambda_{SOC}$ (see supplementary Fig. S3). 
It becomes rational when considering that SOC connects states with the same total angular momentum $J_z=L_z \pm S_z$ as a scattering potential. {When treating SOC perturbatively, } no matter the sign of the SOC, the allowed transitions between the chiral chain and electrode are dictated by non-zero matrix elements of the SOC Hamiltonian ("selection rules"). 
Suppose the orbital polarization in the chiral chain will be in favor of $L_z=+1$ and equal probability for $S_z = \pm \frac{1}{2}$. 
Direct calculation of these matrix elements shows that the allowed transitions are:
\begin{align*}
    \ket{L_z =1,S_z =\frac{1}{2}}_{chiral}\rightarrow & \ket{L_z=1,S_z =\frac{1}{2}}_{lead} \\
    \ket{L_z =1,S_z =-\frac{1}{2}}_{chiral}\rightarrow & \ket{L_z=1,S_z =-\frac{1}{2}}_{lead},\\ 
    & \ket{L_z=0,S_z=\frac{1}{2}}_{lead}
\end{align*}
We can see that these selection rules allow the positive orbital polarization to convert spin states, from down to up, by trading angular momentum between orbital and spin, while an up-down conversion of spins is not allowed. Thus, the positive orbital polarization leads to the positive spin polarization, when scattered by the SOC site. Similarly, a negative orbital polarization generates negative spin polarization. The orbital and spin relation is shown by the same sign between $\sigma_t$ and $L_t$ (also $\sigma_{r'}$ and $L_{r'}$) in Fig.~\ref{fig:orbital-spin}.

\begin{figure}
\includegraphics[width=12cm]{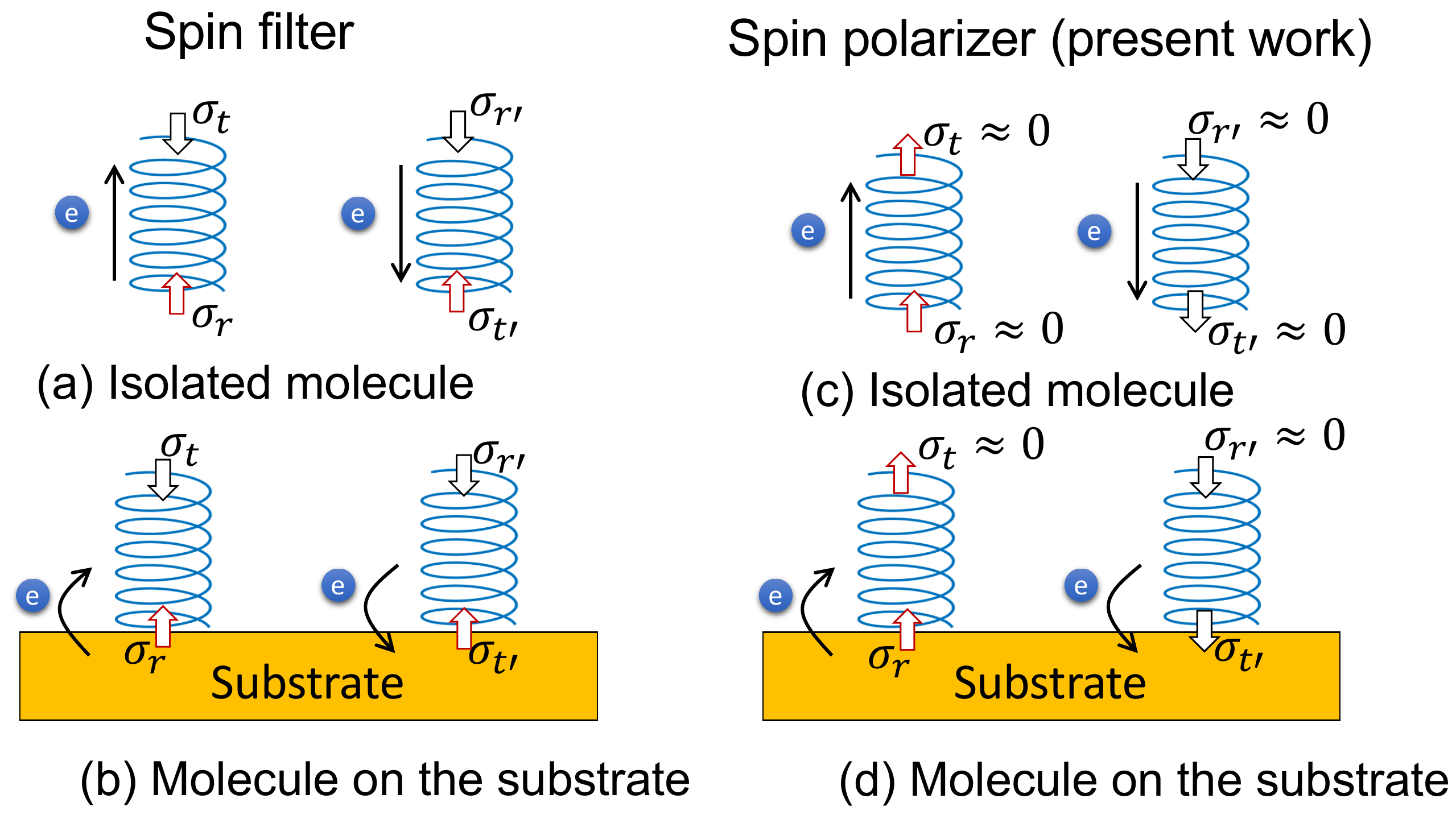}
\caption{\label{fig:transient} 
Transient spin polarization in   (a)\&(c) the isolated chiral molecule and (b)\&(d) a chiral molecule on the surface. The instantaneous charge redistribution is indicated by the black arrow. The spin conductance $\sigma_i$ represents the transient polarization and is defined in the same way as Fig.~\ref{fig:spinfilter}.
}
\end{figure}
\subsection{Discussion on experiments}

As discussed above, the direction of the spin polarization is determined by the current direction and molecule chirality while the magnitude of spin polarization depends on the local SOC.  This observation provides insights on the CISS-induced transient spin polarization of chiral molecules. In chemical reactions or surface adsorption, it is commonly argued in literature\cite{Banerjee2018,Tassinari2019,Naaman2019apl,Ziv2019,Santra2019,ben2017magnetization} that the instantaneous charge displacement leads to opposite spin polarization at both ends of a chiral molecule (as illustrated in Fig.~\ref{fig:transient}a-b) by assuming the spin filter scenario. 
Following the picture of spin polarizer, we expect that both ends exhibit the same sign of spin polarization (see Fig.~\ref{fig:transient}c). If the chiral molecule is isolated far from a strong SOC region (e.g., a substrate or electrode), the magnitude of spin polarization may be negligible because of weak SOC.
If the molecule is close to a heavy metal surface (Fig.~\ref{fig:transient}d), the interface region may develop substantial spin polarization. Additionally, such transient spin polarization may vanish soon after the spin lifetime when the system approaches equilibrium. 

It is noteworthy that the spin polarizer mechanism gets more rational when one considers the previous 
AHE experiments \cite{smolinsky2019electric,mishra2020b} . When the top gate ejects/extracts  electrons through a layer of chiral molecules into/out of doped-GaAs or GaN, 
the magnetization was induced in the doped semiconducting layer and monitored by the AHE.
Switching the gate voltage was found to reverse the sign of induced magnetization in the semiconductor.
The spin polarizer naturally indicates opposite spin polarizations in the semiconductor after reversing the tunnelling direction for the given chirality (Fig.~\ref{fig:spinfilter}d-e). 

This is consistent with the experimental observation that the AHE changes sign after reversing the gate voltage over the chiral molecule layer\cite{smolinsky2019electric,mishra2020b} . 
In contrast, the spin filter scenario would indicate no sign change in the semiconductor side (Fig.~\ref{fig:spinfilter}a-b) after switching the gate voltage. 

\begin{figure}
\includegraphics[width=12cm]{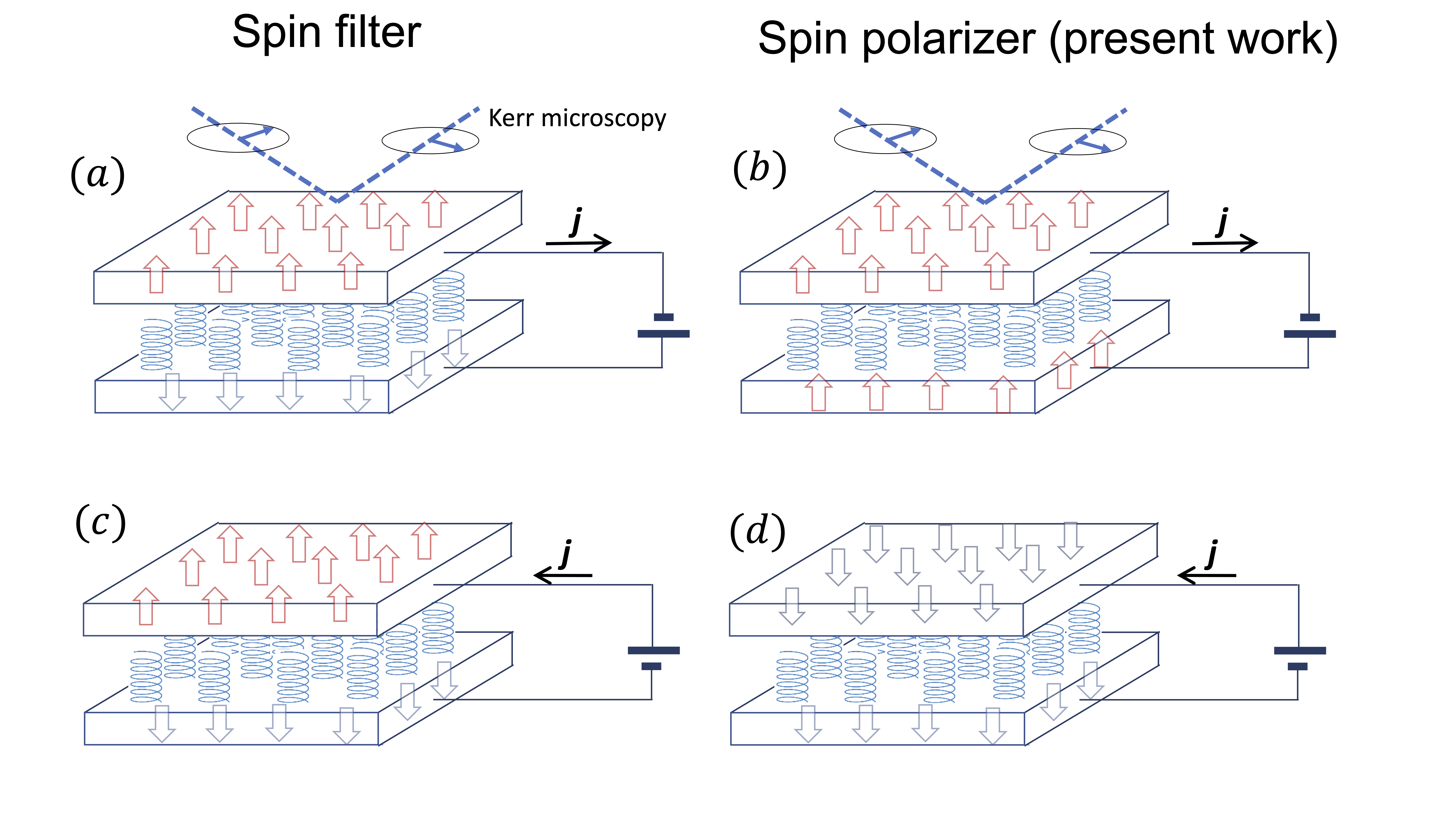}
\caption{\label{fig:new experiment} 
Schematics of a two-terminal device where two nonmagnetic electrodes sandwich a chiral molecule thin film. The current-induced magnetization (large red and blue arrows) on the top electrode can be measured with a scanning SQUID or Kerr microscopy.  (a)\&(c) Spin polarization in a spin filter scenario where the current-induced magnetization remains the same after reversing the current ($\mathbf{j}$). (b)\&(d) Spin polarization in a spin polarizer scenario where the current-induced magnetization switches sign after reversing the current.
}
\end{figure}

We propose that the induced spin polarization in electrodes can also be detected by spectroscopies that are sensitive to surface magnetization, e.g., the magneto-optical Kerr effect \cite{qiu2000surface} or the scanning SQUID\cite{vasyukov2013scanning,persky2022studying} . 
Figure \ref{fig:new experiment} depicts the schematic for a two-terminal CISS device with nonmagnetic electrodes sandwiching a chiral film. 
As current is driven through the system, both electrodes get spin polarized. The induced-magnetization on the top electrode can be detected by a Kerr microscope or scanning SQUID. 
The magnetization detected this way would remain with the same sign for both current flow directions in the spin filter scenario but would change sign according to the spin polarizer mechanism, unambiguously discriminating the two mechanisms. In this experiment, the amplitude of the magnetization will be proportional to the current density and and also the magnitude of SOC in the top electrode. 

\section{Conclusions}
In summary, we proposed that the spin polarizer model is more rational than the spin filter for the chiral molecule in the CISS device. 
Here, the transmitted and reflected electrons exhibit the same sign in spin polarization as a leading order effect. 
This scenario provides deeper understanding on the CISS-driven spin polarization and alternative explanations on the induced AHE and transient spin polarization. The spin polarizer (filter) leads to opposite (same) spin polarization direction in a given electrode when reversing the current direction. {Thus, the current-direction-specific spin polarization provides a smoking-gun evidence to verify the spin polarizer effect in experiments.}

\section{Methods}

A more general version of the calculation of spin conductance using the scattering matrix can be found in part I of the supplementary information, where the the relation between transmitted and reflected conductance are derived for a general observable.

All the conductance calculations were performed with Kwant\cite{Kwant} . Parameters of the model and band structure (supplementary Fig. S2) of leads and chiral chain can be found in part II of the supplementary information.

\begin{acknowledgement}
The authors thank helpful discussions with Ron Naaman. B.Y. acknowledges the financial support by the European Research Council (ERC Consolidator Grant ``NonlinearTopo'', No. 815869) and the MINERVA Stiftung with the funds from the BMBF of the Federal Republic of Germany. N.P. was supported by the NRF of Korea government (MSIT)(No. NRF-2019R1A2C2089332). 



\end{acknowledgement}

\providecommand{\latin}[1]{#1}
\providecommand*\mcitethebibliography{\thebibliography}
\csname @ifundefined\endcsname{endmcitethebibliography}
  {\let\endmcitethebibliography\endthebibliography}{}


\section{Supplementary Information}
\begin{enumerate}
\item Section I. {Derivation of Polarization Equality Between Transmitted and Reflected States.}
\item Section II. {Parameters of Model and Band Structure.}
\item Figure S1. Reflection of an electron by a chiral molecule when the electron spin is not favored by CISS. 
\item Figure S2. Band structure of the chiral chain and linear chain.
\item Figure S3. Spin polarization for negative SOC on both sides.
\end{enumerate}


\section*{Supplementary material}

# Unusual Spin Polarization in the Chirality Induced Spin Selectivity

Yotam Wolf,[†] Yizhou Liu,[†] Jiewen Xiao,[†] Noejung Park,[‡] and Binghai Yan[*,†]

[†]Department of Condensed Matter Physics, Weizmann Institute of Science, Rehovot 7610001, Israel

[‡]Department of Physics, Ulsan National Institute of Science and Technology (UNIST), Ulsan, 44919 Republic of Korea

E-mail: binghai.yan@weizmann.ac.il

# Supplementary information

## I  Derivation of Polarization Equality Between Transmitted and Reflected States

Here, we shall show that in a two terminal system with no dissipation and leads that respect time reversal symmetry, the polarization of transmitted and reflected currents with respect to a general traceless operator, $\mathcal{O}$, is the same. An incoming or outgoing state from the left or right has $2N$ components, corresponding to $N$ orbitals, each 2-fold degenerate in spin.

In the density matrix formalism, the relation between an incoming state and an outgoing state is given by:

$$\rho_{out} = S\rho_{in}S^\dagger = \begin{pmatrix} r & t' \\ t & r' \end{pmatrix} \begin{pmatrix} \rho_L^{in} & 0 \\ 0 & \rho_R^{in} \end{pmatrix} \begin{pmatrix} r^\dagger & t^\dagger \\ t'^\dagger & r'^\dagger \end{pmatrix} \tag{S1}$$

1

Carrying out the matrix multiplication yields the outgoing states in the left and right:

$$\begin{pmatrix} \rho_L^{out} & * \\ * & \rho_R^{out} \end{pmatrix} = \begin{pmatrix} r\rho_L^{in}r^\dagger + t'\rho_R^{in}t'^\dagger & * \\ * & r'\rho_R^{in}r'^\dagger + t\rho_L^{in}t^\dagger \end{pmatrix} \tag{S2}$$

When current is incoming only from the left, $\rho_R^{in} = 0_{2N \times 2N}$. In addition, under the assumption of an incoherent sum of modes incoming from the left, $\rho_L^{in} = \mathbb{1}_{2N \times 2N}$. Plugging into the above equation:

$$\rho_L^{out} = rr^\dagger \equiv \rho_r \qquad \rho_R^{out} = tt^\dagger \equiv \rho_t \tag{S3}$$

Similarly, when current is incoming only from the right:

$$\rho_R^{out} = r'r'^\dagger \equiv \rho_{r'} \qquad \rho_L^{out} = t't'^\dagger \equiv \rho_{t'} \tag{S4}$$

From the unitarity condition of the density matrix in eq. (4)

$$\rho_r + \rho_{t'} = \mathbb{1} \tag{S5}$$

Multiplying by $\mathcal{O}$ and taking the trace:

$$Tr[\mathcal{O}\rho_r] + Tr[\mathcal{O}\rho_{t'}] = Tr[\mathcal{O}] \tag{S6}$$

As the trace of $\mathcal{O}$ is zero, and using the generalized form of eq. (7) for expectation value, $Tr[\mathcal{O}\rho_i] = \langle \mathcal{O} \rangle_i$, we obtain:

$$\mathcal{O}_r + \mathcal{O}_{t'} = 0 \tag{S7}$$

Similarly, using the second unitary relation in eq. (5), we obtain:

$$\mathcal{O}_{r'} + \mathcal{O}_t = 0 \tag{S8}$$

When these outcomes are combined with the selectivity inherent to the $C_2$ symmetric system, $\mathcal{O}_t = -\mathcal{O}_{t'}$, we obtain:

$$\mathcal{O}_r = \mathcal{O}_t = -\mathcal{O}_{t'} = -\mathcal{O}_{r'} \tag{S9}$$

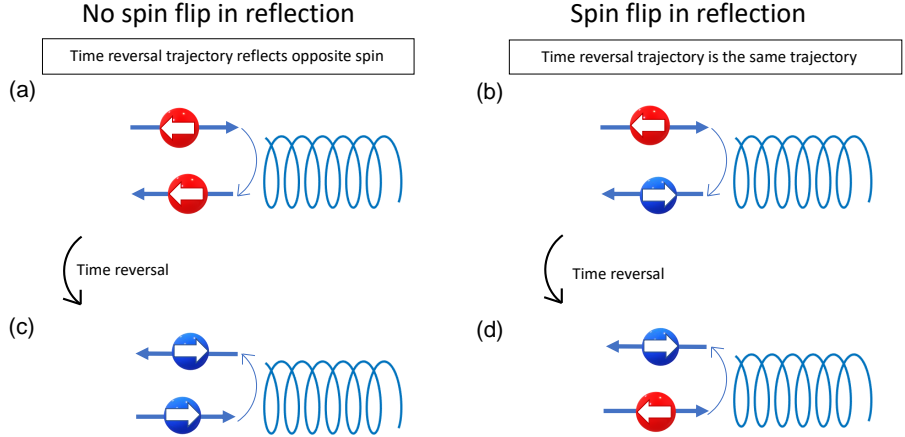

Figure S1: Reflection of an electron by a chiral molecule in front of a non-magnetic lead when the electron spin is not favored by CISS. Since the system is non-magnetic, the time-reversed trajectory (reversing both spin direction and velocity direction) of the electron should preserve the original spin selectivity. (a)&(c) For a spin filter, the reflection involves no spin flip, i.e., the reflected electron has the same spin as the incoming electron. (c) is the time-reversed trajectory of (a). However, (c) indicates the originally favored spin in (a) is reflected (becomes unfavorable). (b)&(d) For a spin polarizer, the reflection involves a spin flip process, i.e., the reflected electron has the opposite spin as the incoming electron. (b) and its time-reversal trajectory (d) are essentially the same process without contradiction, where the same spin gets reflected (remains unfavorable). In summary, the spin selectivity changes as reversing time in (a)&(c) while it remains the same in (b)&(d).

## II  Parameters of Model and Band Structure

In the model we work with electrons in the basis of the of the 3 $p$ orbitals and the up and down states of spin half: $\{|p_x\rangle, |p_y\rangle, |p_z\rangle\} \otimes \{|\uparrow\rangle |\downarrow\rangle\}$. The parameters will all be given in units of $eV$. In addition, they will be shown as an outer product of $3 \times 3$ matrices in the $p$ orbital basis, with $2 \times 2$ matrices in the spinor basis.

In the left and right lead, we use the nearest neighbor hopping matrix:

$$t_{lead} = \begin{pmatrix} -2.4 & 0 & 0 \\ 0 & -2.4 & 0 \\ 0 & 0 & 1.6 \end{pmatrix} \otimes \mathbb{1}_{2\times 2} \tag{S10}$$

The on-site matrix term of the left and right lead is set to zero. The band structure of the leads can be seen in Fig. (S2b).

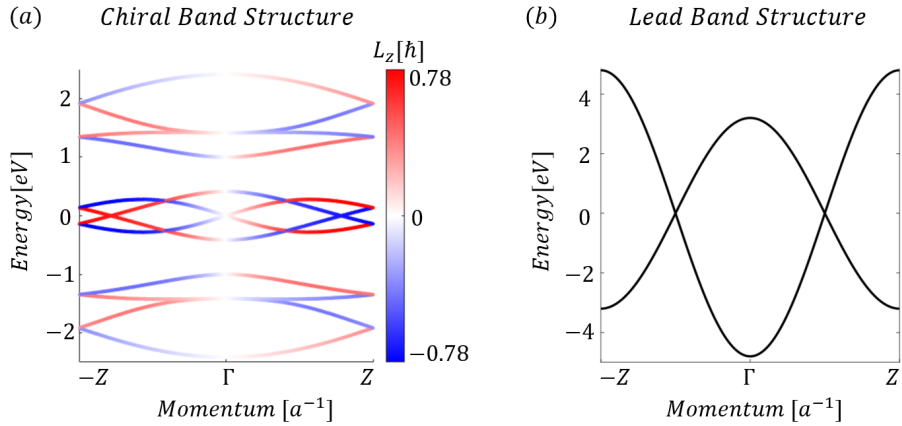

Figure S2: (a) Band structure of chiral molecule in the infinite chain form. The blue and red colors represent the orbital projection. (b) Band structure of the lead.

The interface between the left lead and the chiral molecule contains one atom. The hopping from the lead to this atom is given by the above hopping matrix (S10), and the hopping into the chiral molecule is given by the following hopping matrix:

$$t_{SOC} = \begin{pmatrix} -1.5 & 0 & 0 \\ 0 & -1.5 & 0 \\ 0 & 0 & 1 \end{pmatrix} \otimes \mathbb{1}_{2\times 2} \tag{S11}$$

The onsite term of the atom in the interface is the SOC Hamiltonian:

$$h_{SOC} = \lambda_{SOC,L} \sum_{i=x,y,z} l_i \otimes \sigma_i \tag{S12}$$

4

Where $\sigma_i$ are the three Pauli matrices and $l_i$ are the (unitless) angular momentum operators as $3 \times 3$ matrices in the $p$ orbital basis. $\lambda_{SOC,L}$ denotes the spin-orbit coupling constant, $L$ indicates that this is the interface with the left lead.

For the interface between the chiral molecule and the right lead, we use the exact same model, we only change the spin-orbit coupling constant $\lambda_{SOC,R}$.

The hopping in the chiral molecule is nearest neighbors and is given by the following matrix:

$$t_{chiral,ij} = \begin{pmatrix} t_{xx} & t_{xy} & t_{xz} \\ t_{yx} & t_{yy} & t_{yz} \\ t_{zx} & t_{zy} & t_{zz} \end{pmatrix} \otimes \mathbb{1}_{2\times 2} \quad (S13)$$

$$t_{xx,ij} = t_\pi sin^2\phi_{ij} + cos^2\phi_{ij}(t_\sigma sin^2\theta_{ij} + t_\pi cos^2\theta_{ij})$$
$$t_{yy,ij} = t_\pi cos^2\phi_{ij} + sin^2\phi_{ij}(t_\sigma sin^2\theta_{ij} + t_\pi cos^2\theta_{ij})$$
$$t_{zz,ij} = t_\sigma cos^2\theta_{ij} + t_\pi sin^2\theta_{ij}$$
$$t_{xy,ij} = t_{yx} = sin\phi_{ij}cos\phi_{ij}(t_\sigma sin^2\theta_{ij} - t_\pi cos^2\theta_{ij})$$
$$t_{xz,ij} = t_{zx} = cos\phi_{ij}sin\theta_{ij}cos\theta_{ij}(t_\sigma - t_\pi)$$
$$t_{yz,ij} = t_{zy} = sin\phi_{ij}sin\theta_{ij}cos\theta_{ij}(t_\sigma - t_\pi)$$

where $\theta_{ij}$ and $\phi_{ij}$ are the spherical coordinates of site $j$ with respect to site $i$. In our model, we use a helix with a $C_4$-skew symmetry, such that $\phi_{ij}$ can adopt the angles $\frac{\pi}{8}, \frac{\pi}{8}+\frac{\pi}{2}, \frac{\pi}{8}+\frac{2\pi}{2}, \frac{\pi}{8}+\frac{3\pi}{2}$ and $\theta_{ij} = \pm\frac{\pi}{4}$. In the model, the chiral molecule contains 8 atoms and $t_\pi = -0.5eV$, $t_\sigma = 1.5eV$. The onsite terms in the chiral molecule are set to zero. The sign of $\phi_{ij}$ indicates the chirality of the chain. If chirality is reversed, we only need flip the sign of $\phi_{ij}$ in Eq. S13. The band structure of the chiral chain can be seen in Fig. (S2a).

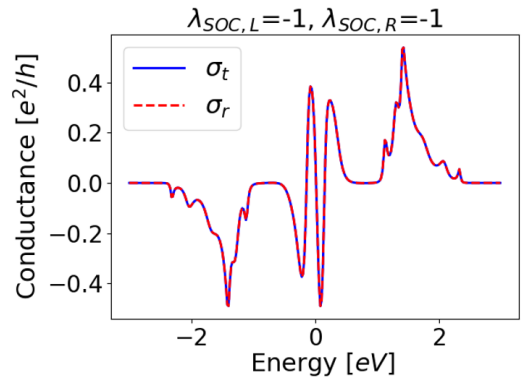

Figure S3: Spin polarization of $t$ and $r$ for negative SOC on both sides. Same sign of polarization as positive SOC.

# Graphical TOC Entry

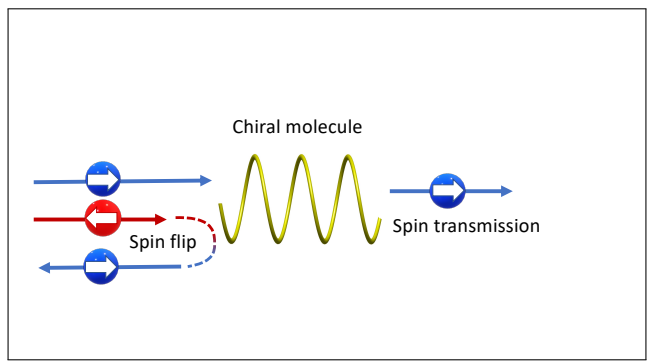







\end{document}